\documentclass[10pt]{amsart}
\usepackage{booktabs}   
\usepackage{array}      
\usepackage{graphicx}   
\usepackage{longtable}  
\usepackage{caption} 
\usepackage{graphicx}    
\usepackage{amssymb,amsmath,amsthm} 
\usepackage{xcolor}      
\usepackage{hyperref}    
\usepackage{lipsum}      

\usepackage[a4paper,margin=1in]{geometry} 
\title{Unveiling temperature and phase boundaries in laser-driven shocked and released copper: insights from ultra-fast X-ray Absorption Spectroscopy up to 300 GPa}

\author{\textsuperscript{1}S. Balugani,\textsuperscript{1}J.A. Hernandez}
\address{\textsuperscript{1}European Synchrotron Radiation Facility, Grenoble, France}

\author{\textsuperscript{2,}\textsuperscript{3}F. Brieuc, \textsuperscript{2,}\textsuperscript{3}J. Boust}
\address{\textsuperscript{2}CEA, DAM, DIF, F-91297 Arpajon, France}
\address{\textsuperscript{3}Université Paris-Saclay, CEA, LMCE, F-91680 Bruyères-le-Châtel, France}

\author{\textsuperscript{4,}\textsuperscript{5}P. Hesselbach}
\address{\textsuperscript{4}GSI Helmholtzzentrum für Schwerionenforschung GmbH, Planckstraße 1, 64291 Darmstadt, Germany.}
\address{\textsuperscript{5}Institut für Angewandte Physik, Goethe-Universität Frankfurt am Main, Max-von-Laue Str. 1, 60438 Frankfurt, Germany}
\author{\textsuperscript{1}N.Sévelin-Radiguet}
\author{\textsuperscript{2,}\textsuperscript{3} V. Recoules}
\author{\textsuperscript{1} O.Mathon}
\author{\textsuperscript{6} D.E.Eakins}
\address{\textsuperscript{6} Department of Engineering Science, University of Oxford, Parks Road, Oxford, OX1 3PJ, UK}
\author{\textsuperscript{7} H.Doyle}
\address{\textsuperscript{7} First Light Fusion, Yarnton, England, United Kingdom}
\author{\textsuperscript{8} A.Ravasio}
\address{\textsuperscript{8}LULI, CNRS, CEA, Sorbonne Université, École Polytechnique – Institut Polytechnique de Paris, Palaiseau, France}
\author{\textsuperscript{1} R.Torchio}

\begin{document}
\maketitle
\begin{abstract}
Cu is an ubiquitous material used in industry for its great thermal and electrical properties.
Studying the high-pressure high temperature properties of copper (Cu) is relevant for nuclear fusion research as projectiles and flyers used in hypervelocity impacts are generally made of copper, where it is used also in the design of the nuclear fusion targets. Recently, a solid (fcc)-solid (bcc) phase transition has been detected in shock compressed Cu with X-Ray Diffraction. Here, we present a study on shock compressed copper up to 300~GPa and 7100~K probed by single pulse (100~ps FWHM) X-ray Absorption Spectroscopy (XAS). Based on the analysis of the XAS spectra, we provide structural identification and bulk temperature measurements along the Hugoniot up to the melting. The collection of XAS spectra under release conditions, i.e. at later times than the breakout time of the shock wave, helped constraining the experimental fcc-bcc and solid-liquid phase boundaries. In particular, we report the first bulk temperature measurement in shock compressed copper on the melting plateau located between 237(40)~GPa and 5750(1130)~K and 261 (27) GPa and 6240 (1155) K and on liquid copper at 300 GPa and 7100 K.
\end{abstract}

\section{Introduction}
Shock compression experiments have gained significant attention in recent decades due to their ability to reach extreme pressure and temperature conditions in materials. Recently, dynamic compression techniques, such as high-power lasers and gas guns, have been combined with advanced X-ray sources, like synchrotrons \cite{1}\cite{2}, \cite{3} and X-ray free-electron lasers (XFELs) \cite{4},\cite{5} \cite{6},\cite{7} revolutionizing this field by enabling the study of materials' microscopic structures under extreme conditions using X-Ray Diffraction (XRD), X-ray Absorption Spectroscopy (XAS) and X-Ray Imaging (XRI). 
Copper is commonly used as a pressure standard in high-pressure research, benefiting from well-established equations of state  and its presumed lack of solid-solid phase transitions. As a result, copper has played a crucial role in calibrating other high-pressure standards, such as ruby and gold \cite{8}. Known for its excellent electrical and thermal conductivity, copper has been widely used for technological applications,. In recent years, copper has been increasingly studied because of its applications in Inertial Confinement Fusion research, where it is used as a flyer and impactor in hypervelocity impact experiments \cite{9}\cite{10}  and as well it is included in the complex target designs of fusion research. 
Understanding copper's behavior under high pressure and temperature is essential for advancing knowledge in material science, metallurgy and fusion science. For a long time, copper was believed to show a monomorphic solid (fcc) in a large P,T range. Indeed, static compression studies \cite{11} up to 153 GPa and ambient temperature show a stable fcc phase. It was only in 2014 when a theoretical work \cite{12} first showed that the bcc phase in nano-polycrystalline Cu is the most energetically favored phase between 100 GPa and 200 GPa when calculating the shock loading using Molecular Dynamics simulations. Then, a theoretical study based on atomistic simulations \cite{13}  found the fcc → bct (body-centered tetragonal) transition at around 80 GPa in single-crystal bulk Cu using both classical and ab initio Molecular Dynamics simulating the shock compression.  Since then, a large experimental effort has been devoted in looking for this transition upon shock loading. The first experimental study that has reported a phase transition from the fcc to the bcc phase upon shock compression at pressures around 180 GPa used XRD as probe \cite{14}. Here the phase change was attributed to the formation of stacking faults, later supported by a theoretical work on Molecular Dynamic simulations \cite{15}. A second work, based on laser-driven shock probed by XRD and MD calculations, confirmed the transition as due to a Pitch-distortion mechanism and underlined that the bcc phase can only be stabilized at high temperature \cite{16}. On the other hand, the fcc-bcc transition is not observed in ramp compression experiments, shockless events resulting in high pressure but low temperature, even up to TPa range \cite{14}\cite{17}.
The fcc-bcc copper's phase transition has primarily been studied under only single shock compression conditions, i.e. along the Hugoniot curve, whereas extended fcc-bcc phase boundaries over a large P,T range have been only theoretically estimated \cite{16},\cite{18},\cite{19}.
Additionally, the Cu melting curve remains largely unexplored at high P,T, probably due to the challenges posed by its high reflectivity, which inhibits effective laser absorption in the LH-DAC setup. Only one experimental study has utilized LH-DAC to melt copper up to 100 GPa, using the laser speckle method to detect the melting  \cite{16}, though without any X-ray microscopic probe. This study was used to anchor the theoretical melting curve calculated by \cite{17}.
Finally, uncertainties remain regarding copper's principal Hugoniot, which represents the set of endstates in the pressure-temperature (P-T) phase diagram that a material can achieve under shock compression starting from initial conditions of ambient density and temperature. Different predictions for the melting plateau exist, with SESAME table 3336 \cite{18} and neural-network-based Density Functional Theory (DFT) calculations \cite{19} suggesting ranges of 230–300 GPa and 300–350 GPa, respectively. In previous works \cite{14}, \cite{20} it was observed the complete melting respectively at 275.6 GPa and 280 GPa upon single shock loading probed by XRD. Sound velocity measurements have detected a kink around 185 GPa- possibly due to the solid-solid phase transition- and a second one between 232 GPa and 265 GPa – possibly indication of melting \cite{21}. \\
In the present work, we conducted a laser-driven shock compression experiment on copper, using ultra-fast X-ray Absorption Spectroscopy (XAS) at the copper K-edge, performed at the High-Power Laser Facility at ID24 energy-dispersive beamline at the European Synchrotron Radiation Facility (ESRF), France\cite{1}. The pressure was determined using two Velocity Interferometer System for Any Reflector (VISARs). Temperature measurement in shock compression experiments is notoriously difficult for opaque materials, with techniques like Streaked Optical Pyrometry relying on surface thermal self-emission, only measurable above 3000 K. Here, we employed the Extended X-ray Absorption Fine Structure (EXAFS) part of the XAS signal, which starts 30 eV above the absorption edge, coupled to MD based simulations, to derive the bulk temperature of copper. The EXAFS oscillations are sensitive to the average ionic temperature through the Debye-Waller factor, which accounts for atomic displacement around equilibrium positions influenced by both thermal and structural disorder.
Most importantly, our XAS data also allow to detect the shock-induced fcc-bcc transition and subsequent melting not only upon single shock but also during release process - the adiabatic unloading event that decreases both pressure and temperature – allowing us to set new and extended constrains on the fcc-bcc phase boundary and melting curve of copper.

\section{Main Results}
\subsection*{XAS data on shocked copper, structural modifications upon shock loading}
\begin{figure}[!h]
    \centering
    \includegraphics[width=0.6\linewidth]{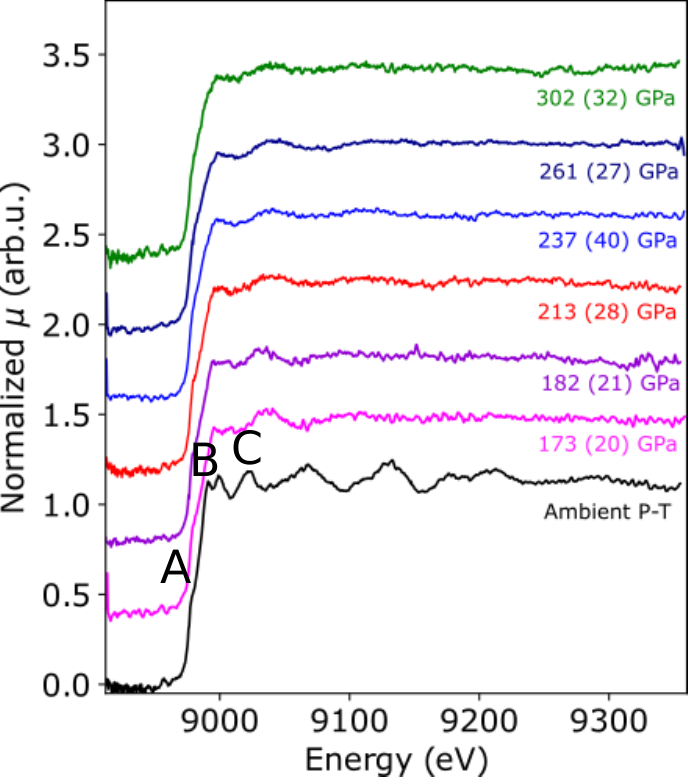}
    \caption{\textbf{K-edge XAS spectra of shock compressed copper along the Hugoniot curve.}The K-edge XAS spectra correspond to spectra of shock compressed copper from ambient conditions of pressure and temperature (black curve) up to 302 GPa (green curve) along the Hugoniot. The spectra are shifted vertically according to the increasing pressure.}
    \label{fig:xas cu exp}
\end{figure}

\begin{figure}[!h]
    \centering
    \includegraphics[width=0.5\linewidth]{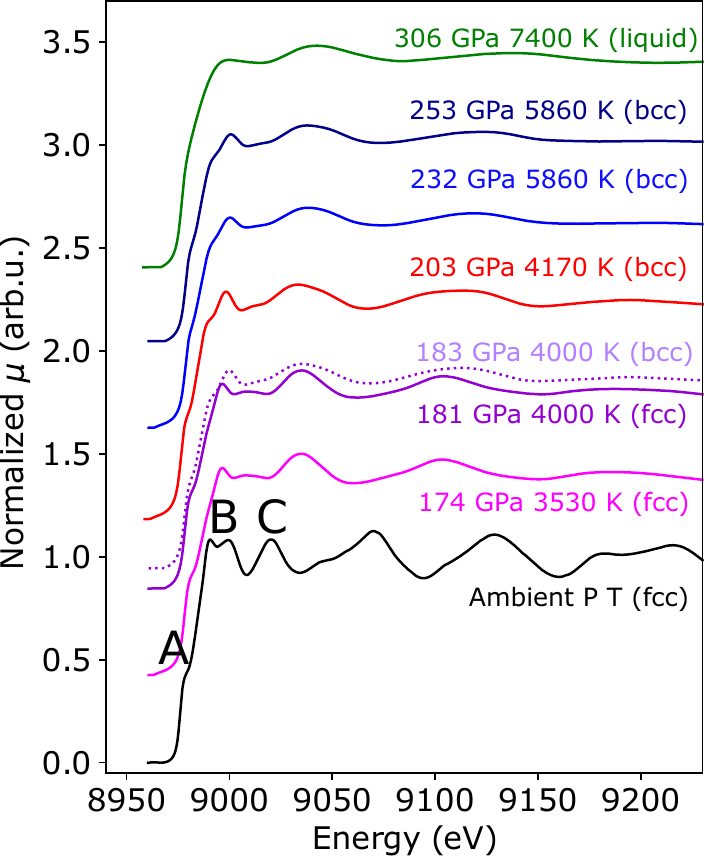}
    \caption{\textbf{Simulated K-edge XAS spectra of shock compressed copper along the Hugoniot curve. }The spectra were generated averaging over four DFT-MD snapshots separated by at least 500 fs and considering each atom of the simulation box as the absorber each time.The dotted spectrum is simulated at the same conditions of the spectrum illustrated with the solid line and same color tonality, but in different crystalline phases.}
    \label{fig:xas simu}
\end{figure}

\begin{figure}[!h]
    \centering
    \includegraphics[width=0.8\linewidth]{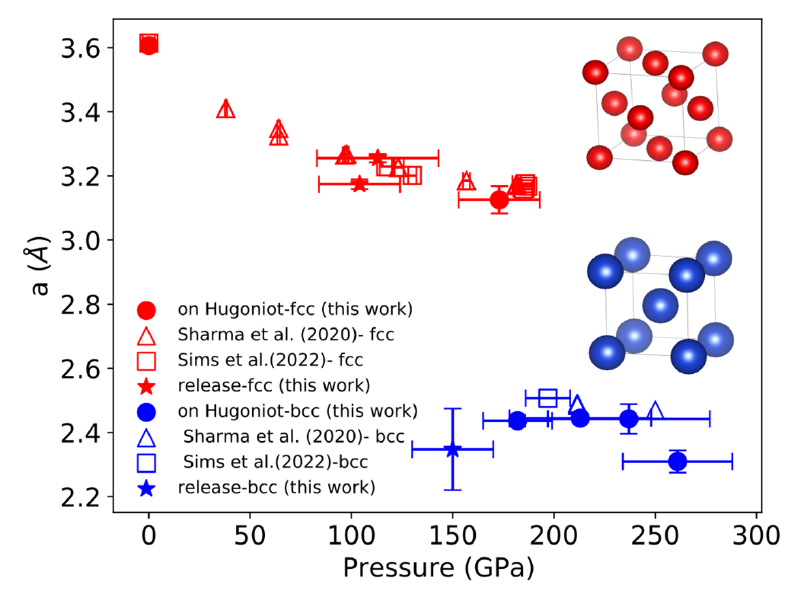}
    \caption{\textbf{First-shell coordination distance as a function of pressure in copper.} The fitting of the first EXAFS oscillation has been carried out using the Artemis software in a crystalline environment determined from the phase determination presented above. The observed drop at around 180 GPa in the first shell distance a is a further confirmation of the phases. Experimental data from previous works are also provided for comparison \cite{14},\cite{20}. A sketch of the unit cell in the fcc (red) and bcc (blue) environment is illustrated.}
    \label{fig:second shell copper}
\end{figure}

Figure \ref{fig:xas cu exp} shows the single-bunch (100 ps) X-ray Absorption Spectroscopy (XAS) spectra of shock-compressed copper, arranged in order of increasing compression (P,T conditions). The P- T conditions are estimated as explained in the next paragraph and in the dedicated section in the Supplementary Materiarls (SM). The ambient spectrum shows typical features of Cu fcc phase, with the characteristic double peak shape just after the absorption edge (feature B at around 8990 eV). At increasing applied P,T conditions under shock loading, the EXAFS oscillations are strongly damped, due to thermal disorder and moved to higher energies as the atomic structure is compressed.   Starting from around 180 GPa modification are observed in the shape of  B feature and of the first EXAFS oscillation (C feature, 9020 eV). At more extreme conditions also the edge concavity (A feature, 8976 eV ) gets modified.
XAS simulations based on ab-initio MD are crucial to attribute such changes to structural changes and melting. Following inputs from previous XRD studies\cite{14}\cite{20}, the fcc phase has been simulated up to 180 GPa, the bcc from 180 GPa and the liquid phase at 300 GPa.
The XAS simulations have also proven necessary for determining the crystalline phase of copper at specific pressure-temperature points on its phase diagram. This is because, in the simulations, not only the pressure and temperature-given as input- but also the structural phase are known. Figure \ref{fig:xas simu} shows the XAS simulations of shocked copper. From these simulations, we identified three key descriptors that distinguish between the main crystalline phases: fcc, bcc, and liquid. These descriptors, labeled A, B, and C in Figures \ref{fig:xas cu exp}\ref{fig:xas simu}, can be summarized as: edge concavity, edge peak shape, and the skewness of the first EXAFS oscillation.
The edge concavity serves as a distinct signature for the liquid phase, becoming flatter as the material transitions into the molten phase, already outlined in \cite{22},\cite{29}. This feature effectively discriminates between solid and liquid phases. The edge peak shape clearly differentiates the fcc and bcc phases. Specifically, XAS spectra in an fcc-crystal environment display a double peak immediately above the absorption edge, that is replaced by a single peak in the bcc environment. Lastly, we observed a change in the skewness of the first EXAFS oscillation from the fcc to the bcc phases. When two or more descriptors of a phase were present, we associated the sample with a pure phase. If only one descriptor of a phase is identified, we classified it as a mixed phase. The first two descriptors are qualitative, while the second one has been calculated using an in-house python script for the calculation of the skewness, see SM for more details.
Table \ref{table:conditions_copper} summarizes the pressure, temperature and the structural phase of the on-Hugoniot spectra reported in \ref{fig:xas cu exp}. \\ We fitted the EXAFS signals of the shocked copper spectra belonging to the solid range using Artemis software \cite{30} to extract the first coordination shell distance, while imposing the obtained crystalline environment and pressure-temperature conditions. Figure \ref{fig:second shell copper} shows the unit cell distances- mathematically derived from the fit of the first shell distances- as a function of pressure, compared to previous XRD shock data from \cite{14} \cite{20}. Here, also release state data are provided, but a discussion about them follows in the next section. The observed drop in the unit cell distance around 180 GPa further supports the occurrence of a phase transition from fcc to bcc, thereby validating our method using the three descriptors.
\\ To summarize this part, we observed that copper remained in a pure fcc phase up to 173 (20) GPa and 3765(660) K, after which a transition to the bcc phase occurred. \\ Two points, corresponding to a mixed solid-liquid copper phase at 237(40) GPa and 5750 (1130) K and 261(27) GPa and 6240 (1155) K, are consistent with previous doppler velocimetry studies \cite{21}, which identified a melting plateau in this pressure range along the Hugoniot. \\ Finally, we observed liquid copper at 302 (32) GPa. \\

\begin{table}[!h]
\centering
\caption{Experimental and DFT-MD pressure and temperature data estimated with the impedance matching method for LiF targets and with hydrodynamic simulations (ESTHER) code for diamond targets and with DFT-MD. The average of the obtained values is then associated to the experimental spectra.}
\label{table:conditions_copper}

\resizebox{\textwidth}{!}{%
\begin{tabular}{c c c c c c}
\toprule
Pressure (GPa) & Temperature (K) & Edge concavity & Edge peak shape & Skewness of 1$^{st}$ EXAFS oscillation & Phase \\
\midrule
 173 (20) & 3765 (660) & pronounced & double & positive & fcc \\
182 (21) & 3945 (425) & pronounced & single & positive & fcc + bcc \\
213 (28) & 4965 (930) & pronounced & single & negative & bcc \\
237 (40) & 5750 (1130) & less pronounced & single & negative & bcc + onset melting \\
261 (27) & 6240 (1155) & less pronounced & single & negative & bcc + onset melting \\
302 (32) & 7185 (1305) & flat & single & negative & liquid \\
\bottomrule
\end{tabular}%
}

\end{table}
\subsection*{Determination of the Hugoniot temperature in shocked copper}
\begin{figure}[!h]
    \centering
    \includegraphics[width=0.8\linewidth]{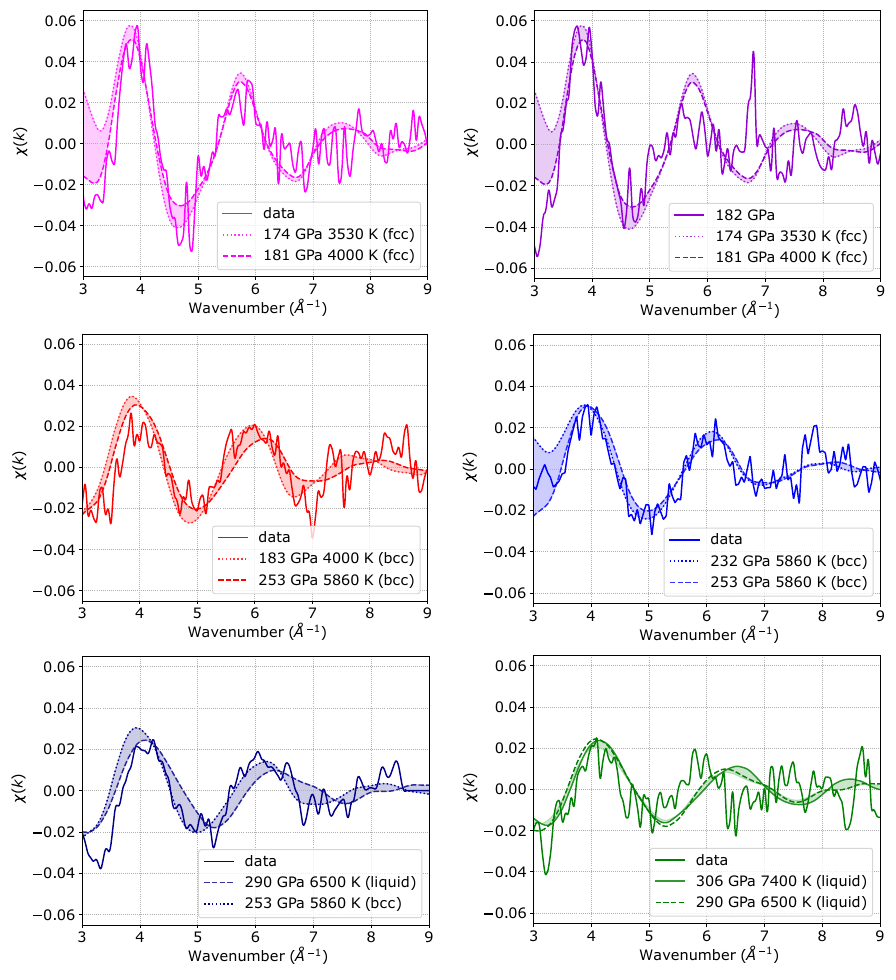}
    \caption{\textbf{Minimization of the intensities of EXAFS $\chi$(k) oscillations for the determination of the Cu bulk temperature.} The solid lines are the experimental data, and the dashed and dotted lines are the simulated ones. Each panel shows the simulated spectra which best minimize the experimental intensities. The experimental intensities were scaled to the simulated ones according to a factor determined at the ambient conditions.}
    \label{fig:minimization cu}
\end{figure}

To measure the shock temperature, we directly analyzed the EXAFS region of the XAS spectra, which exhibits temperature-dependent changes through the Debye-Waller factor \cite{27}. This factor, incorporated into the EXAFS equation, accounts for atomic displacements around their equilibrium positions \cite{28}. \\ In similar studies on iron \cite{22}, \cite{23} and on platinum \cite{24} the temperature was derived by fitting the EXAFS signal. This method requires the knowledge of the Debye temperature as a function of compression, which depends on the Gruneisen parameter. Since the Gruneisen parameter for copper is not well characterized \cite{25}, also the Debye temperature is not well known. Therefore, we used an alternative method previously applied to ramp-compressed copper \cite{26}, in which simulated XAS spectra are generated using Density-Functional Molecular Dynamics (DFT-MD) at given pressure and temperature. A least-squares minimization approach is then employed to identify the simulated spectrum that best matches the experimental data. This method allows fo Our method follows that of \cite{26}, where we minimized the differences between the experimental and simulated EXAFS intensities using the Least Squares Minimization approach. This procedure is more advantageous as it allows the simultaneous simulation of structural disorder and temperature effects. 
The simulated spectra were generated from FEFF calculations and DFT-MD trajectories as described in the Materials and methods section. To ensure consistency, we scaled the intensities of the simulated signal to match the experimental data at ambient conditions, using the experimental ambient signal as a reference. This scaling factor was then fixed during the intensity minimization of the shocked spectra. The scaling procedure also accounts for the experimental resolution. To the best of our knowledge, this is the first successful application of this method to principal Hugoniot shock data for deriving the temperature of shocked copper. \\ Figure \ref{fig:minimization cu}  shows the EXAFS $\chi$(k) signal for both the experimental data and the best-matched simulated spectra. Additionally, we calculated the expected temperature from the hydrodynamic ESTHER simulations, and the final temperature of copper was determined as the average of the minimization-derived temperature and the ESTHER simulation output.
This method provides an additional method to derive the pressure of shocked copper and the final pressure values presented in Figure \ref{fig:xas cu exp}  are the average between the impedance mismatching (LiF window targets) or hybdrid method (diamond window target) and the minimization method.
\subsection*{Off-Hugoniot states}
\begin{figure}[!h]
    \centering
    \includegraphics[width=1\linewidth]{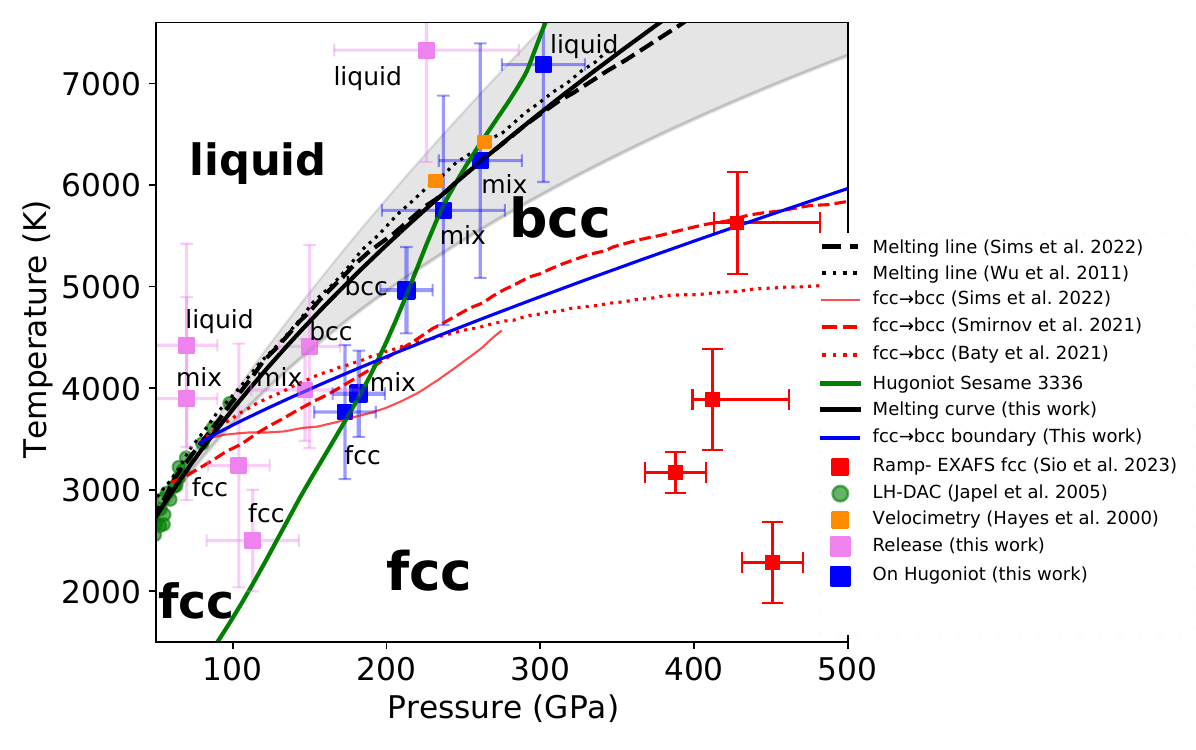}
    \caption{\textbf{Pressure-temperature phase diagram of copper.}The measured points are depicted in blue (on Hugoniot) and in violet (off-Hugoniot) squares and their crystal phases are indicated. The melting lines from \cite{20} and from \cite{17} are reported with a dashed and a dotted black line, respectively. The fcc-bcc phase boundary previously determined by \cite{39}, \cite{31} and by \cite{20} are illustrated with dashed, dotted and solid thin red lines. The here derived fcc-bcc boundary is shown with a thick blue line. The on-Hugoniot shots from \cite{21} at the onset melting and in the molten phase are illustrated with the orange squares. Ramp compressed data of fcc copper from \cite{26} are provided in red squares.}
    \label{fig:phase diagram all points}
\end{figure}

In addition to the on-Hugoniot states, this study also investigated the off-Hugoniot states along the release path, measured after the shock wave breaks out from the copper layer. The release wave is an adiabatic process that unloads both pressure and temperature in copper. To probe a release state, the X-rays must be synchronized with the drive laser to capture data after the shock wave has broken out from the copper. Probing these release states expands the accessible range of pressure and temperature conditions, offering a broader exploration of the phase diagram, which is useful for phase boundary determination. 
To select the appropriate X-ray probing time, we used hydrodynamic simulations to ensure that we probed at a time where uniform pressure and temperature conditions had been achieved. In fact, during the release process, there are times when the sample exists across a wide range of pressure and temperature values due to the interaction of multiple release waves generated at the interfaces, and probing at these moments could result in XAS data that span a too large region of the phase diagram.
The thermodynamic conditions probed for the release states have been determined from hydrodynamic ESTHER simulations, once that they have been scaled such to reproduce the shock timings measured by the VISAR. \\  Figure \ref{fig:phase diagram all points} shows all the experimental points covered in this study, both on Hugoniot (in blue) and off Hugoniot (in pink).
Using the method of the three descriptors validated for the on-Hugoniot states, we associated the structural phase to each of the off Hugoniot states. The collection of on- and off-Hugoniot states has opened the access to the exploration of the phase boundaries: the solid-liquid phase boundary and the fcc-bcc phase boundary.

\subsubsection*{\textbf{Experimental fcc-bcc phase boundary in copper}}
The fcc-bcc phase boundary has been derived by fitting a dataset composed of the on- and off-Hugoniot points collected in this work at 104 GPa and 3239 K (fcc), 173 GPa and 3765 K (fcc), 147 GPa and 3982 K (mix), 182 GPa and 3945 K (mix) and 213 GPa and 4965 K (bcc). The fitting equation was taken from \cite{31} and reads:
\begin{equation}
    T(P)=T_{ref}+\frac{dT}{dP}(P-P_0)-a(P-P_0)^b
    \label{fcc bcc phase boundary}
\end{equation}
The fixed and fitted parameter of the model are summarized in Table \ref{tab:fit_params}.
\begin{table}[!h]
    \centering
    \caption{\textbf{Fitting parameters for the fcc-bcc phase boundary determination.} The $T_{ref}$, $dT/dP$ and $P_0$ have been considered from \cite{31}. a and b have been fitted to Equation \ref{fcc bcc phase boundary}. }
    \label{tab:fit_params}
    \begin{tabular}{c c c c c}
        \toprule
        $T_{\text{ref}}$ (K) & $dT/dP$ (K/GPa) & $P_0$ (GPa) & $a$ & $b$ \\
        \midrule
        3460 (fixed) & 19.9 (fixed) & 79 (fixed) & 10 (4) & 1.06 (0.06) \\
        \bottomrule
    \end{tabular}
\end{table} \\
The derived phase boundary is displayed by the blue solid curve in Figure \ref{fig:phase diagram all points}.
\subsubsection*{\textbf{Melting curve of copper}}
The melting curve of copper has been derived by fitting the Simon-Glatzel equation, which reads:
\begin{equation}
    T=T_{ref} \left(1+\frac{P-P_0}{a}\right)^{1/c}
\end{equation}
with $P_0$ set to ambient condition and $T_{ref}$ to 1358.5 K. The fitted parameters gave a=18(6) and c=1.8(0.2). The error at 1$\sigma$ on the melting curve was determined using the errors from a MonteCarlo loop iterated 200 times over 20 repetitions. The obtained melting curve and the related uncertainty is provided in Figure \ref{fig:phase diagram all points}. The dataset to fit included the previous experimental work on LH-DAC by \cite{16} up to 100 GPa and three points from this work: the two points acquired on the Hugoniot at 237 (40) GPa and 5750 (1130) K and at  261 (21) GPa and 6240 (1155) K belonging to the melting plateau, and one point from the release path at 70 (20) GPa and 3897 (1000) K. This last point has been interpreted as mixed solid liquid because differently from its close point at a slightly higher temperature at 4422 K ($\approx$500 K difference) it shows a more pronounced edge, typical of the solid phase.  The point at 70 GPa and 4422 K is in the liquid phase as it is in great agreement with the DFT-MD simulation we carried out in the liquid phase at ambient pressure and 4500 K, as illustrated in the SM. 

\section{Discussion}
In this work, we have derived for the first time, to the best of our knowledge, the bulk temperature of shock-compressed copper along its principal Hugoniot using a novel method -pioneered for EXAFS data on solid copper upon ramp compression by \cite{26}- that minimizes the intensities of experimental and DFT-MD generated EXAFS signals up to the liquid region at 302 (32) GPa. This approach allowed us to consider both thermal and structural disorders affecting the Debye-Waller factor, which is responsible for dampening EXAFS oscillations. Our method addressed the notorious challenges in temperature measurements of compressed copper, which are difficult in both static and dynamic compression due to copper's high reflectivity and opacity, respectively. 
Moreover, we established a qualitative method based on three descriptors in simulated XAS spectra to associate structural phases with shock-compressed XAS data on copper. We observed the fcc-bcc phase transition at 182 (21) GPa and 3945 (425) K upon shock loading. This result agrees well with previous XRD studies \cite{14}, \cite{20}.
The method of three descriptors was then applied to interpret the crystalline phase of XAS spectra collected under the release path. We observed that the fcc-bcc phase boundary is crossed during a release event, allowing us to determine the first experimental solid-solid phase boundary in copper. 
Our findings indicate that the copper melting plateau lies between 237 (40) and 5750 (1130) K and 261 (27) GPa and 6240 (1155) K, with liquid copper observed at 302 (32) GPa and 7185 (1305) K. These results align with the SESAME table 3336 and several experimental works, but contrast with theoretical predictions suggesting a higher melting plateau between 300-350 GPa \cite{19}.
Using both on- and off-Hugoniot states, we have derived the first experimental melting curve of copper up to 265 (21) GPa, filling a significant gap in the literature regarding copper's high-pressure behavior. 

\section{Limitations and outlooks}
The temperature determination in this study is associated with large uncertainties. Reducing these uncertainties would require performing DFT-MD XAS simulations on a finer pressure-temperature grid, which, while improving precision, would be a time- and resource-consuming process. However, it is worth noting that the reported errors are likely overestimated. This is supported by the high resolution of the XAS spectra, which enables us to discern fine differences between spectra obtained under close pressure-temperature conditions. This capability suggests that the actual precision of the temperature determination may be better than indicated by the error margins.
Our study provides valuable insights into copper's behavior under extreme conditions, but some limitations remain. We could not investigate the driver of the fcc to bcc phase transition, a topic of ongoing debate in the field. Some studies have identified the formation of stacking faults as the driver \cite{14}, while others suggest temperature as the primary factor \cite{20}. Our data indicates that disorder can impact copper at long ionic distances, as we could only fit the first shell distance but not the second with the Artemis software \cite{30}.
These findings have important implications for our understanding of copper's behavior under extreme pressures and temperatures. The novel method we developed for determining bulk temperature and the experimental melting curve provides important constraints for validating theoretical models and improving our knowledge of material properties in extreme conditions. Future research should focus on resolving the mechanisms driving phase transitions and exploring the effects of disorder on copper's structure at high pressures.
Our study has yet to fully exploit the potential of X-ray Absorption Near Edge Spectroscopy (XANES), which is the region within 50-100 eV above the absorption edge.
This region is particularly rich in information about the electronic structure of materials under shock loading conditions \cite{32}. Interestingly, we have observed contrasting behaviors in the absorption edges of two 3d metals, iron and copper, under comparable pressure and temperature conditions upon shock loading. Iron exhibits a redshift of the absorption edge towards lower energies, while copper displays a blueshift. These findings open an entirely new field of research, and the interpretation of these results is still in progress.

\section{Materials and Methods}

\subsection*{The target}
Two different target designs, which vary only in the rear-side window material, were used. One design employed a LiF window, while the other used a diamond window. The pressure values in copper for the LiF-window targets were determined using the LiF’s particle velocity inferred from the VISAR signal and the impedance mismatching method. For the diamond-window targets, a hybrid approach combining VISAR shock timings, experimental laser pulses, and hydrodynamic simulations was used to estimate the pressure. A detailed explanation of the pressure error determination can also be found in the SM.
The targets consisted of multiple layers: a 50 $\mu$m-thick black Kapton (BK) (Polyimide) ablator layer purchased from GoodFellow, a 2 $\mu$m-thick zinc (Zn) shielding layer to mitigate potential pre-heating from X-rays emitted by the hot plasma ablator, a 6 $\mu$m-thick copper (Cu) layer, and, finally, a 150 $\mu$m-thick lithium fluoride (LiF) window purchased from Asphera or 40 $\mu$m-thick diamond (C) window purchased from Diamond Materials GmbH \& Co. The thicknesses of each layer have been chosen according to hydrodynamic simulations using the ESTHER software\cite{33} to generate uniform pressure and temperature conditions in the Cu layer upon laser-shock loading and to guarantee an optimal signal-to-noise ratio in the acquired XAS spectra. The Zn and Cu layers were directly deposited onto the windows by Nano\&Micro PVD (Etupes, France), while the BK layer was bonded to the Zn layer with a glue layer. A critical aspect in the assembly of targets involving glue is the variability in glue thickness across the surface. For this process, we used a low-viscosity epoxy adhesive, ÅngströmBond® AB9110LV. A small drop of glue was carefully applied to the BK layer using a needle, after which the Zn/Cu/window assembly was affixed to the BK layer. The assembly was left to dry for 24 hours under compression between a plate and a flat rounded tip to ensure uniform glue distribution and prevent the formation of voids. To ensure that the glue layer is homogeneous across targets, we analyzed the breakout times of the shock wave from the copper layer against the laser intensity shown in Fig.S3. We estimated the glue thickness to be approximately 3 $\mu$m.
Pre-heating test to rule out the presence of an unwanted pre-heating of the sample prior the arrival of the shock wave was carried out. This is done by synchronizing the X-ray hundreds of ps after the laser initiation. Also, the initial density of copper has been checked as the initial conditions of the sample need to be perfectly characterized for an accurate determination of the pressure and temperature conditions. These checks are provided in the SM. 
\subsection*{Experimental set-up}
\begin{figure}[!h]
    \centering
    \includegraphics[width=0.9\linewidth]{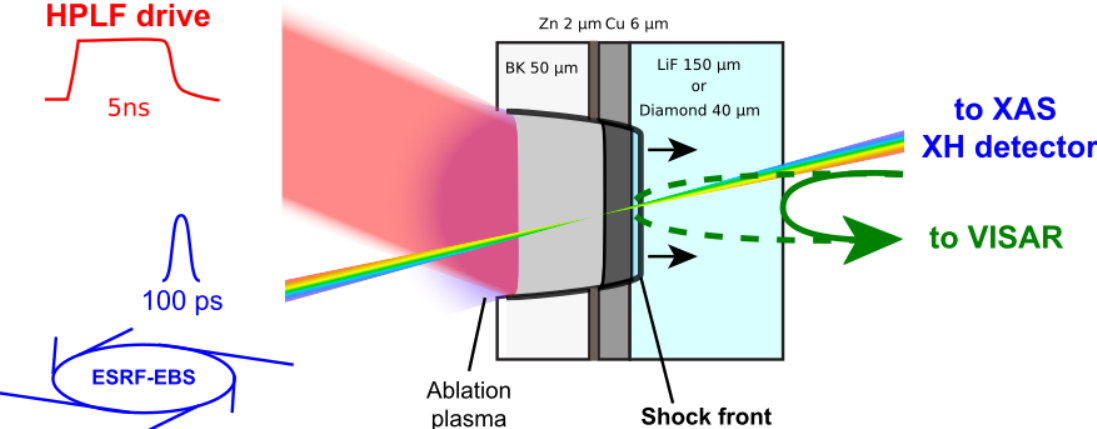}
    \caption{\textbf{Sketch of the experimental set-up.} The HPLF infrared ($\lambda$=1053 nm) laser which drives the shock (in red) and the ESRF-EBS broadband X-rays (in blue) are focused on the target with ±15 degree angle relative to the normal of the target plane. The XAS absorbance spectrum $\mu$L is obtained in transmission by collecting the intensity of a single X-ray through the sample during the shock ($I_1$(E)) and by normalizing it to a reference intensity profile without sample ($I_0$(E)).}
    \label{fig:set up}
\end{figure}

The experiment was conducted at the High-Power Laser Facility (HPLF) at the ID24 energy-dispersive beamline of the European Synchrotron Radiation Facility (ESRF), France \cite{1}. A sketch of the experimental set up is provided in Figure \ref{fig:set up}. Ablation was achieved by focusing 5 ns flat-top pulses at a wavelength of 1053 nm onto a 100 $\mu$m diameter focal spot using a hybrid phase plate. The drive laser energy was varied to explore the Hugoniot of copper (Cu) across a pressure range of 170 to 300 GPa.
Two line-imaging Velocity Interferometer System for Any Reflector (VISAR) operating at $\lambda$=660 nm loaned by GSI Helmholtzzentrum für Schwerionenforschung GmbH (Germany) were used to measure shock timings. For targets with a LiF window, VISARs recorded the apparent particle velocity at the Cu/LiF interface. The pressure in Cu was then determined using the impedance mismatching method between Cu and LiF, assuming the equations of state (EoS) for both materials. The Hugoniot data from \cite{34} was used for Cu, and that from \cite{35} was used for LiF. The error on the pressure here was associated with the MonteCarlo method. A more detailed description of this process is provided in the SM. For targets with a diamond window, VISAR captured shock timings for both the shock wave entering and exiting the diamond window, along with the velocity of the diamond’s free surface. This data was used to calculate pressure values in the Cu layer via ESTHER hydrodynamic simulations \cite{33}, utilizing the SESAME 7830 EoS for diamond (further details are provided in the SM).
Time-resolved single-bunch X-ray absorption spectroscopy (XAS) was performed at the Cu K-edge (8.980 keV) using an energy-dispersive setup in the ESRF’s 16-bunch timing mode. The ID24-ED pink X-ray beam was dispersed and focused onto the sample into a 6 $\mu$m × 6 $\mu$m spot using an elliptically bent Si(111) crystal (polychromator). The beam then diverged and was detected by a 1D position-sensitive detector (XH), allowing the collection of a full XAS spectrum in a single 100 ps[36]. X-ray pulse synchronized with the shock, achieving an energy resolution of 2.2 eV- accounting also for the core-hole lifetime. The delay between the drive laser and the X-ray single bunch was precisely chosen to capture the fully shocked Cu when the shock wave reached the Cu/window interface. Potential pre-heating of the target by X-rays from the coronal plasma was ruled out by analyzing the XAS spectrum at early times, prior to the shock wave reaching the Cu layer as detailed in the SM.
Additionally, XAS spectra of Cu under shock unloading conditions (i.e., off-Hugoniot release states) were measured by synchronizing the X-ray bunch at later times, after the shock breakout from Cu. The thermodynamic conditions of these states were determined using hydrodynamic simulations similarly to the procedure applied to the targets with diamond window and described in the SM. In the simulations, we used the EoS contained in the ESTHER database for Cu and diamond, specifically the SESAME 3336 and SESAME 7270 models. To validate this method against the impedance mismatching one, we compared the Cu and LiF Hugoniots from the SESAME database with those from the experimental works of  \cite{34} and \cite{35}, finding no significant differences in the pressure range probed in this study (see SM).

\subsection*{Theoretical method}
We performed theoretical XAS simulations to compare with the experimental XAS data. Our approach, outlined in \cite{22}, involved simulating the XAS spectra using the FEFF10 \cite{37}software, starting from atomic trajectories derived from Density Functional Theory-Molecular Dynamics (DFT-MD) calculations at specific pressure and temperature conditions. The DFT-MD electronic structure calculations were carried out on a 3 × 3 × 3 Monkhorst-Pack k-point grid, with 2500 electronic bands, ensuring that the eigenstates were converged up to 50 eV above the chemical potential\cite{38}. The Local Density Approximation (LDA) was used for the exchange functional \cite{38}. The PAW dataset included semicore states with 19 outer electrons (3s$\textsuperscript{2}$ 3p$\textsuperscript{6}$ 3d$\textsuperscript{1}0$ 4$\textsuperscript{1}$) and a cutoff radius of 2 bohr \cite{38}. Two projectors were used for the s and p channels, while three projectors were applied for the d channel. The plane-wave cutoff was set to 25 Ha \cite{38}. A grid of points in the density-temperature plane was chosen to simulate atomic trajectories at the measured Hugoniot density, $\rho_H$, as well as at $\rho_H$ ± 0.3 g/cm³, thus accounting for potential experimental uncertainties in density and, consequently, pressure. For the calculation of the absorption spectrum, 4 MD snapshots separated by 500 fs were selected after thermalization. A XAS spectrum was calculated considering each atom of the simulation cell as the absorber each time. We considered cells of 108 atoms (fcc) and of 128 atoms (bcc). The final XAS spectrum was the result of the averaging over the four time-separated snapshots multiplied by 108(128).


\end{document}